\documentclass[twocolumn,epjc3]{svjour3}

\usepackage{graphicx} 
\usepackage[numbers]{natbib}
\usepackage{amssymb}
\usepackage{multirow}
\usepackage{float}

\usepackage{array}

\journalname{Eur. Phys. J. C}

\begin{document}

\newcommand{\simtel}{\textit{sim\_telarray}}

\title{Background Rejection in Atmospheric Cherenkov Telescopes using Recurrent Convolutional Neural Networks}

\author{R.D. Parsons\thanksref{addr1, cor1} \and S. Ohm\thanksref{addr2}}

\thankstext{cor1}{e-mail: daniel.parsons@mpi-hd.mpg.de}
\institute{Max-Planck-Institut f\"ur Kernphysik, P.O. Box 103980, D 69029, Heidelberg, Germany \label{addr1}
           \and
           DESY, D-15738 Zeuthen, Germany \label{addr2}
}

\date{Received: date / Accepted: date}

\maketitle

\begin{abstract}

  In this work, we present a new, high performance algorithm for
  background rejection in imaging atmospheric Cherenkov telescopes. We
  build on the already popular machine-learning techniques used in
  gamma-ray astronomy by the application of the latest techniques in
  machine learning, namely recurrent and convolutional neural
  networks, to the background rejection problem. Use of these
  machine-learning techniques addresses some of the key challenges
  encountered in the currently implemented algorithms and helps to
  significantly increase the background rejection performance at all
  energies.

  We apply these machine learning techniques to the H.E.S.S. telescope
  array, first testing their performance on simulated data and then 
  applying the analysis to two well known gamma-ray
  sources. With real observational data we find significantly improved 
  performance over the current standard methods, with a 20-25\% reduction in 
  the background rate when applying the recurrent neural network analysis.
  Importantly, we also find that the convolutional neural network results are
  strongly dependent on the sky brightness in the source region which has 
  important implications for the future implementation of this method
  in Cherenkov telescope analysis.
    
\end{abstract}

\section*{Introduction}

Historically, one of the largest challenges in ground-based gamma-ray
astronomy with imaging atmospheric Cherenkov telescopes (IACTs) is the
identification and rejection of hadron-initiated air showers based on
shower images. This is due to the extreme outnumbering of gamma-ray
induced air showers by those from cosmic-ray hadrons (a factor $10^4$
in even the brightest fields of view). Therefore, in order to detect
most sources, the difference in development of hadronic and
electromagnetic air showers, which causes a corresponding difference
in the observed IACT camera image, must be used to discriminate
gamma-ray candidates from hadronic background.

Traditionally, this background rejection has been performed through
the use of \emph{Hillas Parameters} \cite{HillasParams}, which
parameterise the cleaned (typically using a two threshold tail cuts
cleaning, e.g. \cite{TailCuts}) camera images using their second
moments. Images from hadronic showers appear to be both, longer and
wider, than those from gamma rays, due to the larger transverse
momentum transfer within the hadronic interactions in the shower
cascade. By placing cuts on the image width and length, the first
generation of very-high-energy (VHE; $0.1$\,TeV$\leq E \leq 50$\,TeV)
gamma-ray sources were detected \cite{CrabWhipple, Mrk421}. In the
following generation of gamma-ray observatories the use of multiple
telescopes to image air showers from different directions improved
their characterisation and the classification capability. The
information from more than one image of the same shower has been combined by using the
Hillas parameters from multiple telescopes to construct mean scaled parameters
\cite{HESSCrab}, for example mean scaled width (MSCW) defined below:
%
%

\begin{equation}
MSCW = \sum_{i=1}^{n} \frac{w_i - \langle w \rangle }{\sigma_{w}} . \frac{1}{n}
\end{equation}

where $w$ is the Hillas width, $\langle w \rangle$ is the expected
width determined from lookup tables derived from Monte Carlo air-shower simulations and $\sigma_{w}$ is the expected RMS of the
width. In this way the Hillas parameters can be combined into a single
parameter with a mean of zero and a standard deviation of
one. Background rejection can then be achieved by placing cuts on both
the MSCW and MSCL parameters. Note that lookup tables are typically
produced for a broad phase-space range, covering a variety of
observing conditions and telescope setups.

The use of Hillas parameters for background rejection has proven
extremely effective in the rejection of hadronic background, however,
it is clear that these parameters do not effectively contain all the
information from individual camera images like asymmetries or
pixel-wise information. By construction,
mean-scaled parameters also average over multiple telescopes employing
different weightings. Necessarily, this leads to a loss of information
on the separation power stored in individual images. Another
limitation of the classical {\it mean-scaled parameter} based box cuts
is that they do not take into account linear and non-linear
correlations between input parameters. Machine-learning techniques
such as random forests \cite{Albert2008}, boosted decision trees
\cite{Ohm2009, Becherini2011} or neural networks \cite{Murach2015}
have been developed and successfully applied to data taken with the
third generation of IACTs based on telescope- and event-wise
input. These algorithms do explore correlations between variables, but
cannot compensate the information lost in the construction of the
input parameters. The same is true for the much more powerful
state-of-the-art likelihood methods that base the classification on
parameters from pixel-wise comparisons between the recorded shower
images and the expected image from a semi-analytical
\cite{deNaurois2009} or template-base model \cite{Parsons2014}.

This paper is organised as following: in the first section we
motivate the usage of deep neural networks to address the apparent
information-loss problem in classical parameter-based IACT classifiers
and introduce the deep neural network designs used for this
study. The following section explains how the networks are constructed and trained, followed by a section addressing the performance
of the network with Monte-Carlo events. Finally,
we will test the performance and stability of the network against real
H.E.S.S. data.

\subsection*{Convolutional Input Layers}

If we wish to move beyond the paradigm of image parameterisation as used in 
state-of-the-art machine-learning techniques such as \cite{OhmTMVA} and \cite{Murach2015}, using a multi-layer perceptron (MLP)
is no longer sufficient.  Although an MLP
could be created using individual image pixels as input, such a
network would require an extremely large training data set in order to
properly classify data. This is due to two effects: firstly the
maximally connected nature of the MLP (each neuron in a layer is
connected to each neuron in the proceeding layer) would create a huge
number of parameters to fix in the case of even a very coarsely
pixelated image. Secondly all spatial information of the pixels
relative to their neighbouring pixels would be lost, therefore the
network would not be stable against the translation of a given
classifying feature through the image.

Convolutional neural networks offer a way around this problem by
instead extracting the information from the shower image in the 
Cherenkov camera itself, and by applying a series of convolutional kernels on it. 
The result of this application is a 2-dimensional feature map of the image. 
Typically in such networks the most important features are selected (and the dimensionality reduced) through the use of a \emph{max pooling} layer, where the maximum value
of the feature map in a given 2D window is selected. The results of
this pooling can then be passed through further convolutional and
pooling layers, allowing features on larger scales than the
convolutional kernel to be extracted. Different CNN architectures have been successfully implemented and applied in particle and astroparticle physics (e.g. \cite{Nature_ML, IceCube_CNN, NOvA_CNN}).

\subsection*{Recurrent Network Layers}

Often in machine learning problems the classification of a number of sequential correlated images (for example images of the same shower seen from different perspectives) is required. Again, in this case the construction of a traditional network structure with each image of the series as an input would introduce an unsatisfactory number of free parameters to the network. Additionally, as the same features are being searched for in all images, such separate input is counterproductive. 

To counter this problem, recurrent layers are constructed in such a way that the correlated inputs can be fed through the same network in sequence with each input modifying the behaviour of the network for all subsequent inputs. In this way the network is able to process inputs while retaining knowledge of information, which has already been seen. Typically, recurrent network implementations such as the long short term memory (LSTM) \cite{LSTM} contain mechanisms to "forget" older inputs, such that the sum of potentially extreme former inputs does not lead to a runaway of the network weights to infinity. One example of a RNN in particle physics, is the work by \cite{ATLAS-Jet-RNN}, where identification of $b$ quarks based on particle jet properties in ATLAS at the LHC is performed. In this work, different tracks associated to the same jet are sequentially input into the RNN, which learns about the correlations between tracks associated to the same vertex. 

The application of a similar convolutional, recurrent network structure have already been demonstrated on IACT simulations \cite{ShilonCNN} with some encouraging results. However, the expected improvements were not seen when applied to H.E.S.S. data. This clearly demonstrates the challenges regarding the stability and reputability when deploying those advanced analysis method on experimental data. Tackling those aspects is one of the primary focus of this work.

\begin{figure}[]
\begin{center}
\includegraphics[width=0.99\columnwidth]{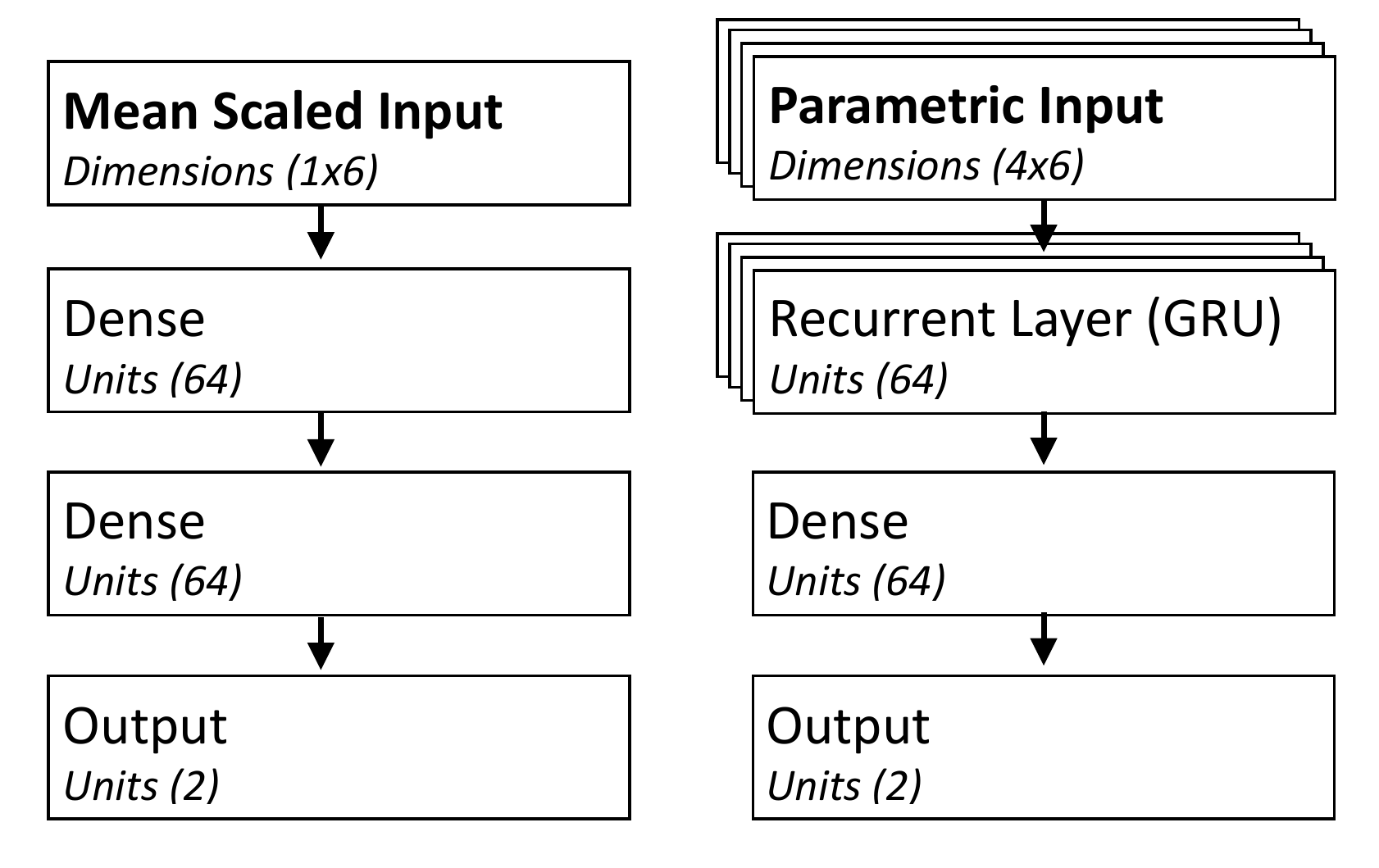}
\caption{Network topology of the Mean Scaled (left) and Hillas-based recurrent (right) networks, stacked boxes show regions of the network where telescope inputs are processed in parallel.}
\label{fig:NetworkTop}
\end{center}
\end{figure}

\begin{figure}[t!]
\begin{center}
\includegraphics[width=0.99\columnwidth]{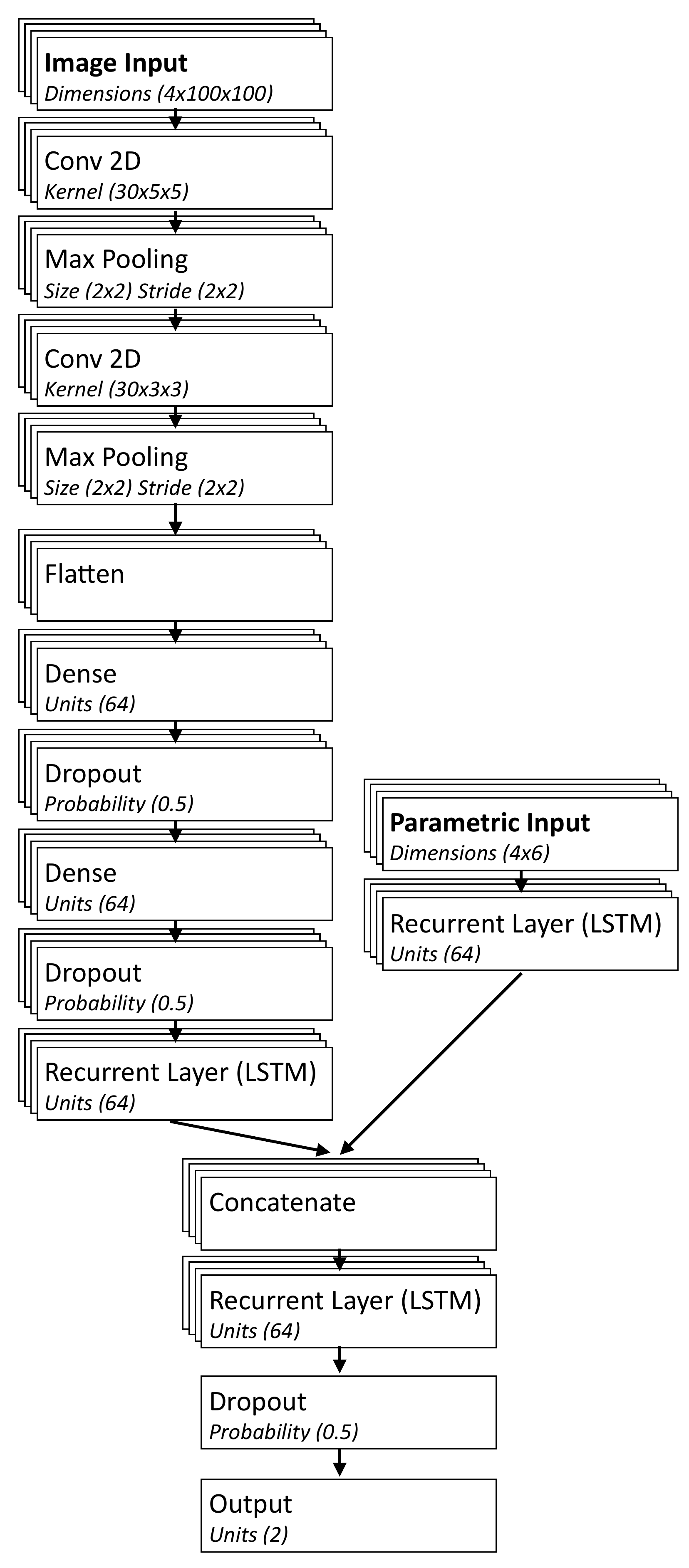}
\caption{Network topology of the convolutional network, stacked boxes show regions of the network where telescope inputs are processed in parallel.}
\label{fig:NetworkTopCNN}
\end{center}
\end{figure}

\section*{Neural Network Design}
\label{sec:network}

In order to quantify the performance of the recurrent neural network on both, simulations and IACT data, three networks were designed using different inputs and network topologies (see figures \ref{fig:NetworkTop} \& \ref{fig:NetworkTopCNN}). All networks were created using the Keras \cite{Keras} python-based machine-learning interface, using \emph{TensorFlow} \cite{TensorFlow} as the back-end module.

\subsection*{Mean Scaled Input}

Firstly a simple multi layer perceptron (MLP) was created using mean scaled parameters as input. This network was created to provide a baseline comparison for the recurrent networks. The input parameters for this network are the mean scaled width and length of the shower (in comparison to both simulated gamma-ray and background events), the reconstructed depth of maximum and the consistency of energy estimates from the different telescopes. These input parameters are similar to those used in the BDT method of \cite{OhmTMVA}, and hence should perform similarly to the technique already implemented in the H.E.S.S. framework. However, a retraining is performed to ensure consistency of the MVA tools used and the training data set. In general this network performs similarly to that described in \cite{OhmTMVA}.

\subsection*{Parametric Input}

The second network created also used the Hillas parameters for the input information, however, rather than combining these parameters using the aforementioned mean scaled method instead the unscaled parameters are used as input and combined within the network using a recurrent layer. The inputs to this network are the Hillas width and length, sum of pixel amplitudes in the cleaned image, reconstructed impact parameter, the displacement of the image centroid from the reconstructed source position and the distance of the image centroid from the camera field of view.

\subsection*{Image Input}

Finally, in order to quantify the effects of adding more image information, a network was created, which also takes the camera images as input. However, as most convolutional algorithms are created to operate on a regular image of square pixels, rather than the hexagonal arrangement used in many IACT cameras (such as those of the H.E.S.S. array) some preprocessing must be performed. Firstly the camera images are cleaned using the standard split level tail cut scheme used in H.E.S.S. \cite{HESSCrab} and then 4 rows of additional pixels are added to the edge of the cleaned image (in the same procedure used by \cite{Parsons2014}). A linear interpolation is then performed between these pixels, using Delauney triangulation, allowing them to be mapped onto a square grid with pixel size of 0.05$^\circ$ and a total width of 5$^\circ$.
Although the image cleaning step is not strictly necessary for the convolutional analysis, the reduction in the number of image pixels greatly increases the speed of the interpolation step and reduced the amount of data stored. Additionally the removal of noisy pixels not in the vicinity of the shower image may help to produce a more stable result. Finally the image is rescaled such that the image intensity lies between 0 and 1 (with negative intensity pixels set to 0). This rescaling was found to greatly ease training and although it does remove some normalisation information from the network the amplitude information is added to the network as part of the parameteric input layer.

Once this preprocessing is complete the data is passed to the convolutional neural network pictured in figure \ref{fig:NetworkTopCNN}. This network takes the interpolated images as input, passing them through two steps of convolution and max pooling and then flattening the resulting feature map into one dimension and passing it through a densely connected neural network. To avoid significant over-training of the network, dropout layers were added to this section \cite{DropoutLayer}. During the training of the network these layers randomly remove a fraction of the network connection (in this case 50\%) to ensure no individual connections can dominate the network. This convolutional section was purposefully designed to be rather simple (in comparison with cutting-edge image classification algorithms) to try to avoid the situation where classification power is dependent on subtle image features present only in simulated data. In this case we sacrifice some potential performance for stability.

The result of this convolutional section is then concatenated with the densely connected layer of the parametric network described earlier and fed into a recurrently layer and ultimately to the output layer. In principle concatenating the results of the network in this fashion is not required to perform image classification however it is useful in this case for two main reasons. Firstly it allows us to assess the rejection power of the information added to the network by the image data over the parametric. Secondly and most crucially it provides information to the network which cannot be easily extracted from the camera images, such as the distance of the telescopes from the shower core. Given a sufficiently large training data set such information could be included the network implicitly, by learning the locations of the telescopes in addition to how to perform event reconstruction. However, this would significantly increase training time and could potentially introduce systematic effects to the results.  

\section*{Network Training}
\label{sec:training}

The three networks were trained using simulated data generated from the CORSIKA Monte-Carlo air shower simulation code \cite{CORSIKA} and the \emph{sim\_telarray} telescope and camera simulation \cite{simtel}. This simulation chain has been proven within the H.E.S.S. and CTA collaborations to provide an accurate representation of the telescope data. To train the network, a sample of simulated gamma rays and protons was created which simulates the performance on the phase 1 H.E.S.S. array (4$\times$12\,m telescopes) at 70\% of their design optical efficiency. Events were simulated in a diffuse cone of opening angle 2.5$^\circ$ with an energy spectrum of E$^{-1.5}$ and an energy range covering from below 100\,GeV to over 100\,TeV (dependent of the simulated species). 
%
%

\begin{figure}[]
\begin{center}
\includegraphics[width=0.99\columnwidth]{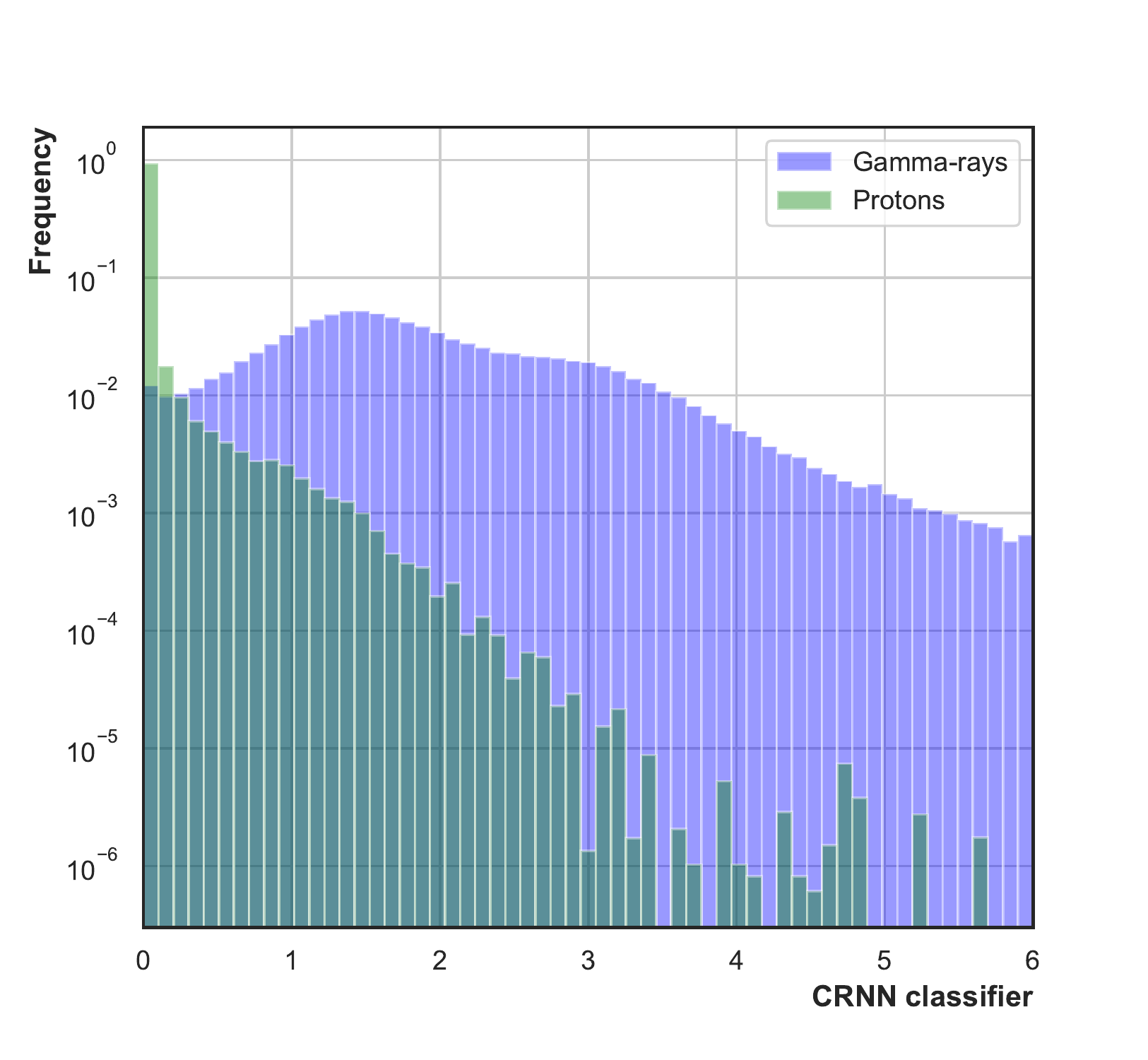}
\caption{Distribution of the CRNN classifier for a sample of gamma-ray and proton events. 
Events are re-weighted to represent an energy spectrum of E$^{-2}$ for both signal and background.}
\label{fig:ClassDist}
\end{center}
\end{figure}

\begin{figure*}[]
\begin{center}
\includegraphics[width=1.0\columnwidth]{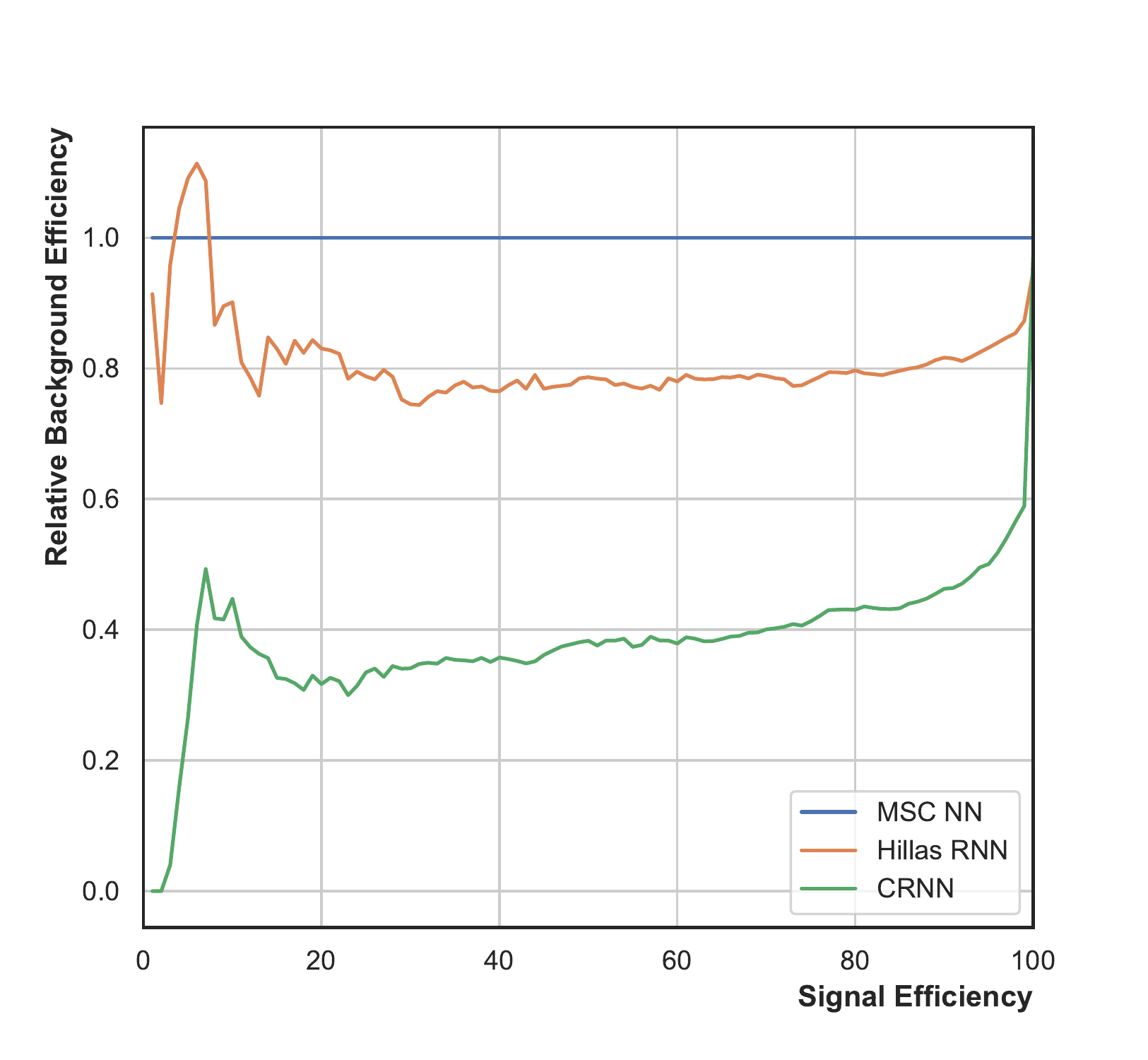}
\includegraphics[width=0.99\columnwidth]{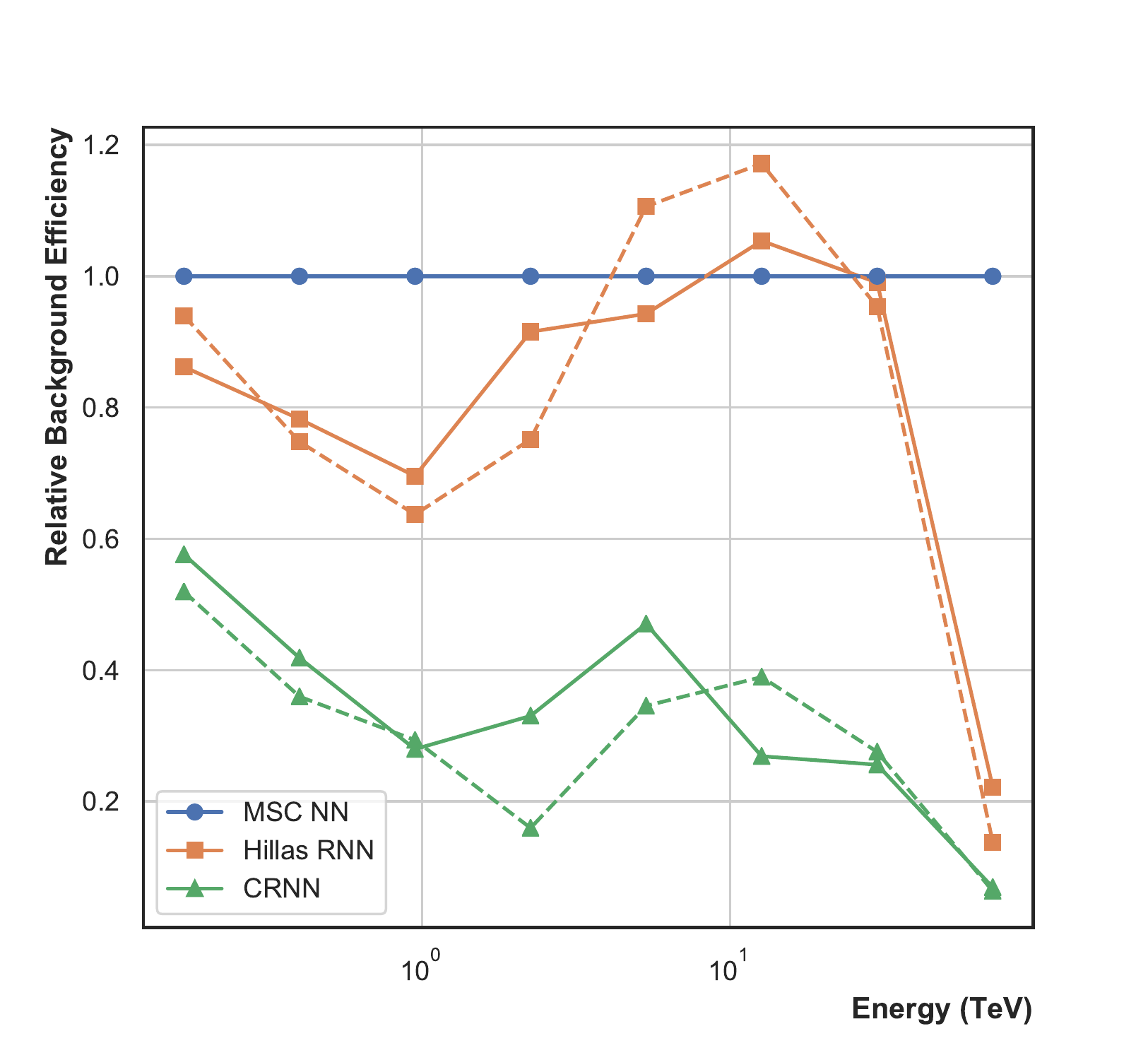}

\caption{Left: Comparison of background rejection performance (as a ratio to mean scaled network performance) vs energy-dependent signal cut for the two recurrent networks. Right: Energy dependence of background rejection performance (as a ratio to mean scaled network performance) for the two recurrent networks at 80\% (solid line) and 60\% signal efficiency.
Events are re-weighted to represent an energy spectrum of E$^{-2}$ for both signal and background.}
\label{fig:ROCFrac}
\end{center}
\end{figure*}

The simulated events were then passed through the H.E.S.S. Analysis Program (HAP) and the standard event selection cuts applied, requiring at least 2 camera images in an event over 60 photoelectrons and with an image centroid less than 2$^\circ$ from the camera centre\cite{OhmTMVA}. The remaining events were then reconstructed using the standard H.E.S.S. Hillas parameter based shower reconstruction and  events reconstructed as lying within the central 1$^\circ$ from the camera centre passed to the neural network. This event selection resulted in a total of around 100,000 gamma-ray and proton events. These events were then split into 4 energy bins (0.1-0.4, 0.4-1, 1-5 \& 5-100\,TeV), with energy ranges chosen as a compromise between keeping a small range to ensure similar events are compared and having sufficient event statistics to perform the training.
The network was then trained in these 4 energy bins using 80\% of the events as the training sample and the remaining 20\% as an independent validation sample used to modify the network learning rate during training.

\section*{Monte Carlo Performance}
\label{sec:perfMC}

Once training was complete the performance of the network was tested using an independent set of Monte-Carlo data representing the four H.E.S.S. phase one telescopes at 70\% of their nominal optical efficiency and a zenith angle of 20$^\circ$. In order to represent the typical data taking mode of H.E.S.S., gamma-ray events were simulated as a point source with an offset from the telescope pointing direction of 0.5$^\circ$, while protons were simulated as a diffuse source with an opening angle of 2.5$^\circ$, however only events reconstructed in the the central 1$^\circ$ were included in performance evaluations. 

The output of the neural network when evaluated on this dataset is a classification value between 0 and 1, roughly representing the probability that the event is a gamma ray ($P_{\gamma}$). However, as most gamma-ray events lie so close to 1 this classifier was reformulated to make the distribution more easily visible.

\begin{equation}
\zeta = - \,\log_{10} (1 - P_{\gamma})
\end{equation}

The resultant classifier distribution of $\zeta$ is shown in figure \ref{fig:ClassDist} and is strongly peaked at 0 for the tested protons and lies between 0 and 10 for gamma-ray events. However, it is typically useful when cutting on this parameter to select events based on an energy-dependent (as the classifier distribution is typically strongly energy-dependent) gamma-ray efficiency.

\subsection*{Background Rejection Performance}

Figure \ref{fig:ROCFrac} (left) shows the  performance of the recurrent networks in comparison with the traditional mean scaled parameter based network at different gamma-ray efficiency cut levels. The performance improvement of the recurrent networks is clear, with improvements seen at all levels of signal efficiency. The Hillas RNN shows around a 20\% reduction in background in comparison to the mean scaled network, while the CRNN shows almost a 60\% improvement in rejection power.

Figure \ref{fig:ROCFrac} (right) shows the energy dependent comparison of background rate to mean scaled network at performance at 80\% and 60\% gamma-ray efficiency.
In the lower energy bins ($<$5\, TeV) a clear improvement is seen in the performance of the recurrent networks over the mean scaled network. A $\sim$20-25\% reduction is seen in the proton rate in the Hillas RNN at both 80\% and 60\% signal efficiency . Such an improvement at low energies could be expected due to the relatively large fluctuations in the Cherenkov light distribution in this energy range, potentially resulting in significantly different images being seen in the different telescopes. In this case taking the mean of the shower parameters will result in a loss of information and performance, whereas the recurrent network can use the full information from all telescopes.

The CRNN shows an even larger improvement in the lowest energy bins, showing a reduction proton rate of more than 60\% at both 80\% and 60\% signal efficiency. At low energies the convolutional layers are able to provide additional image information to the background rejection, most likely using information from pixels that were eliminated from the Hillas parameter construction by the image cleaning.

Above 5\,TeV, the Hillas RNN performance matches closely the mean scaled network at both signal efficiencies, as the more well defined air showers in this energies range reduce the observed differences in the different telescopes. The larger images available in this energy range, however, provide significant information to the CRNN maintaining and in some cases improving on the 60\% improvement in background rejection seen at lower energies.


\label{sec:perfReal}

\subsection*{Sensitivity to Night Sky Background Level}
\label{sec:perfNSB}

\begin{figure}[]
\begin{center}
\includegraphics[width=0.99\columnwidth]{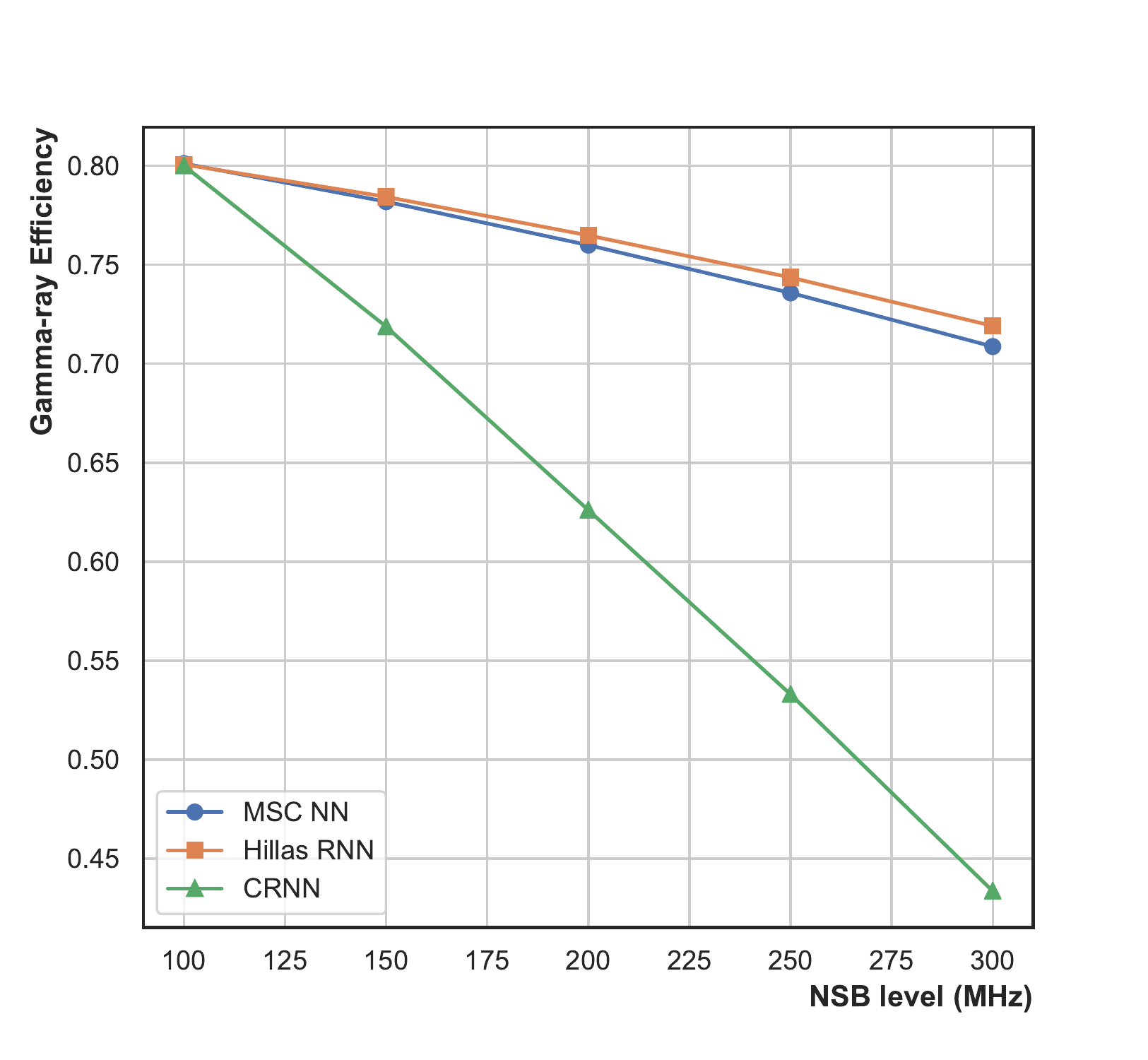}

\caption{Degradation in gamma-ray acceptance for the three tested networks as a function of nightsky background level.}
\label{fig:NSB}
\end{center}
\end{figure}

The performance of the neural networks presented so far were evaluated at the nominal, per pixel, NSB of 100\,MHz used within H.E.S.S. simulations. However, this simulated value is a compromise between that observed  level in extragalactic regions of as low as 50\,MHz and that seen in the Galactic plane, which can reach to 300\,MHz or above in some bright regions (e.g. the Carinae region). In order to test the robustness of the networks against differing levels of noise we created gamma-ray simulations at 5 different NSB levels (100-300\,MHz) and tested the fraction of events passing a background rejection cut defined using the 100\,MHz simulations.

\begin{figure*}[]
\begin{center}

\includegraphics[width=0.948\columnwidth]{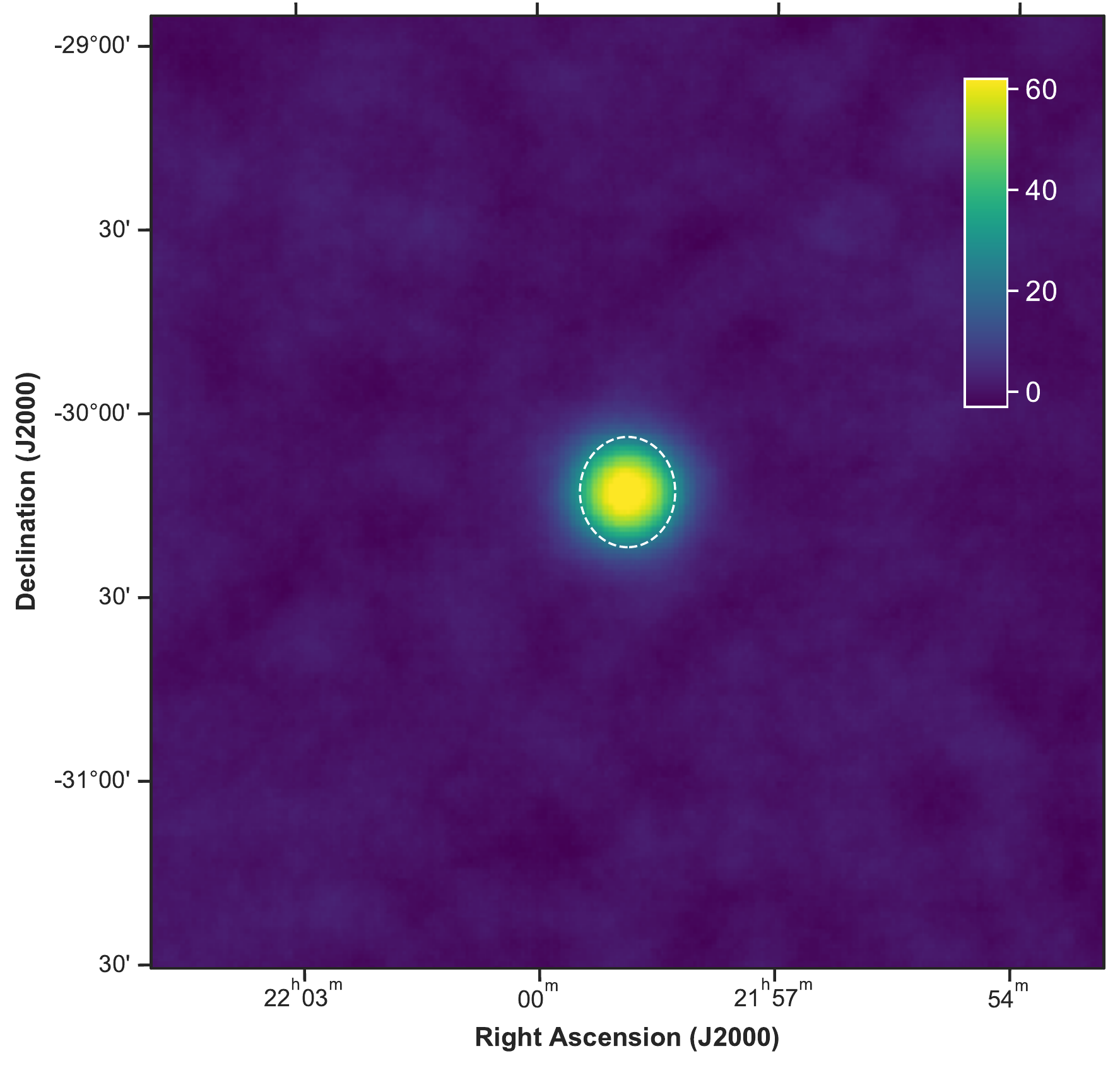}
\includegraphics[width=1.04\columnwidth]{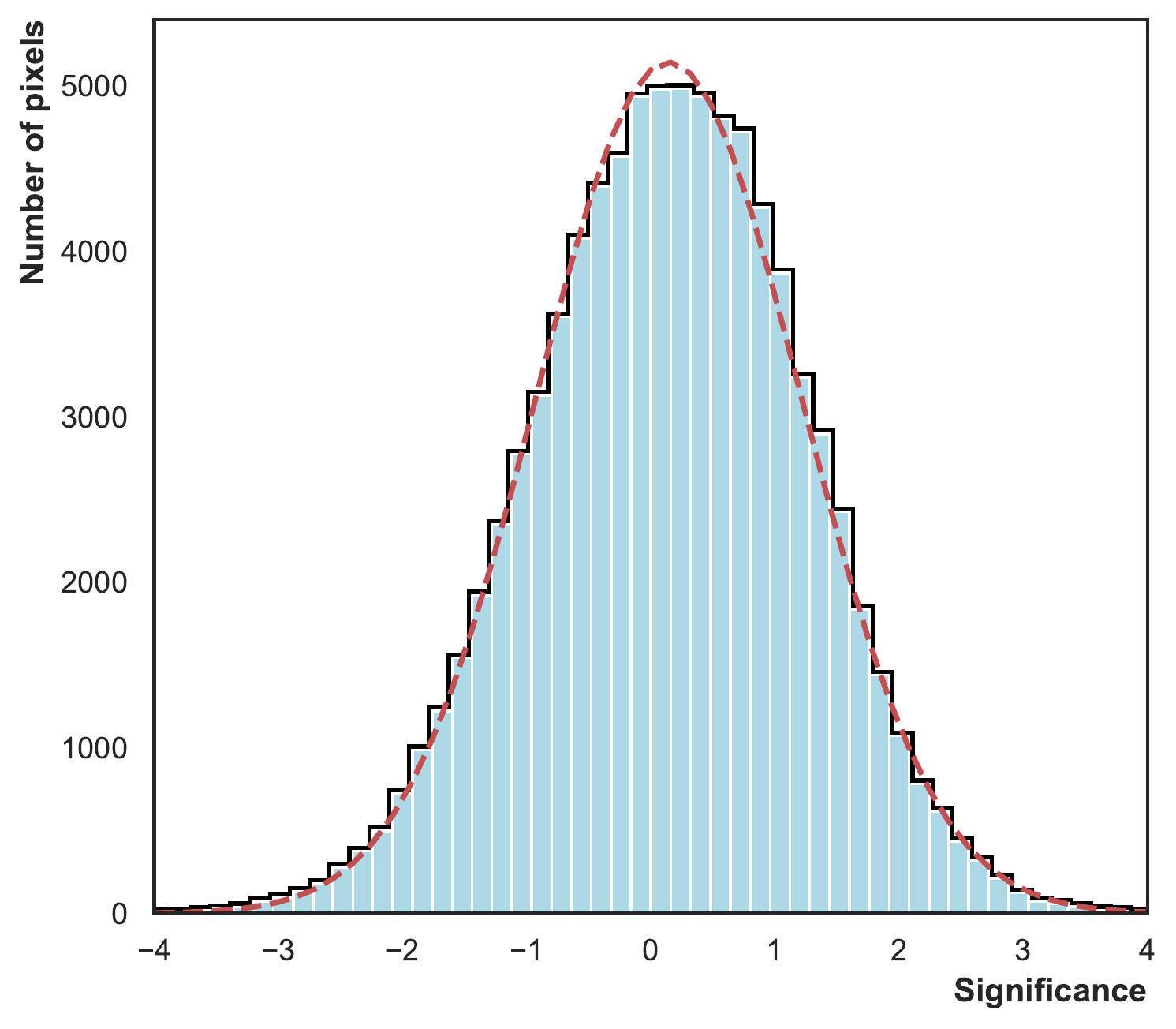}

\caption{Left: Significance map of the PKS 2155-304 region created using an oversampling radius of 0.12$^\circ$ at 80\% gamma-ray efficiency, the position of the source is marked with the dotted circle. Right: 1D distribution of significance from signal-free pixels (solid histogram), shown in comparison with the best-fit Gaussian.}
\label{fig:SignificanceMap}
\end{center}
\end{figure*}

Figure \ref{fig:NSB} shows the acceptance of gamma-ray events at the different NSB levels, when defining the background rejection cut level at 80\% gamma-ray acceptance based on the simulations at 100\,MHz NSB. It is clear from these results that the Hillas-based networks are rather robust, falling in acceptance by only around 10\% from 100 to 300 MHz.
The robustness of the Hillas parameters can be understood from the two tail cut cleaning levels (typically 5 \& 10\,p.e.) applied to the image being significantly higher than the expected pixel to pixel fluctuations resulting from NSB noise (around 1\,p.e. at 100\,MHz in H.E.S.S.), resulting in the noise level in the included pixels being low.
The CRNN however is strongly affected by the NSB, with a 50\% reduction in gamma-ray event acceptance from 100 to 300\,MHz.
This reduction in acceptance is due to the convolutional portion of the network using the uncleaned image sections and therefore include pixels which contain no signal and only nightsky background noise. Therefore increasing the noise level in the non-signal distribution clearly affects the CRNN classifier distribution, producing lower $\zeta$ values for equivalent events.

This strong sensitivity to NSB is clearly a concern when evaluating the performance of the CRNN on data and it must be ensured that the results are compared with simulations of an appropriate NSB when extracting results.

\section*{Performance on H.E.S.S. Data} \label{perfReal}

Tests on an independent Monte Carlo have shown a significant increase in performance of both recurrent neural networks over the mean scaled network. However, these simulations are based on an idealised representation of the instrument behaviour. In reality camera images may contain a number of issues that affect the quality of the data, for example some camera pixel may be broken or the level of night sky background may vary across the field of view. The  network was therefore tested on H.E.S.S. phase one data. The outcome of this analysis was then used to test the stability of the results and check the performance in comparisons to the predictions of Monte Carlo simulations.

In order to ensure the different classifiers are compared in a fair way, energy dependent cut sets were created for the three classifiers that maintain a fixed efficiency of gamma rays passing cuts. In this case values of 80\% and 60\% gamma-ray efficiency were chosen as typical values for soft and hard cuts respectively. Table \ref{tab:stats} shows the statistics for the cuts tested using the three different neural network configurations.

\begin{figure*}[]
\begin{center}
\includegraphics[width=0.99\textwidth]{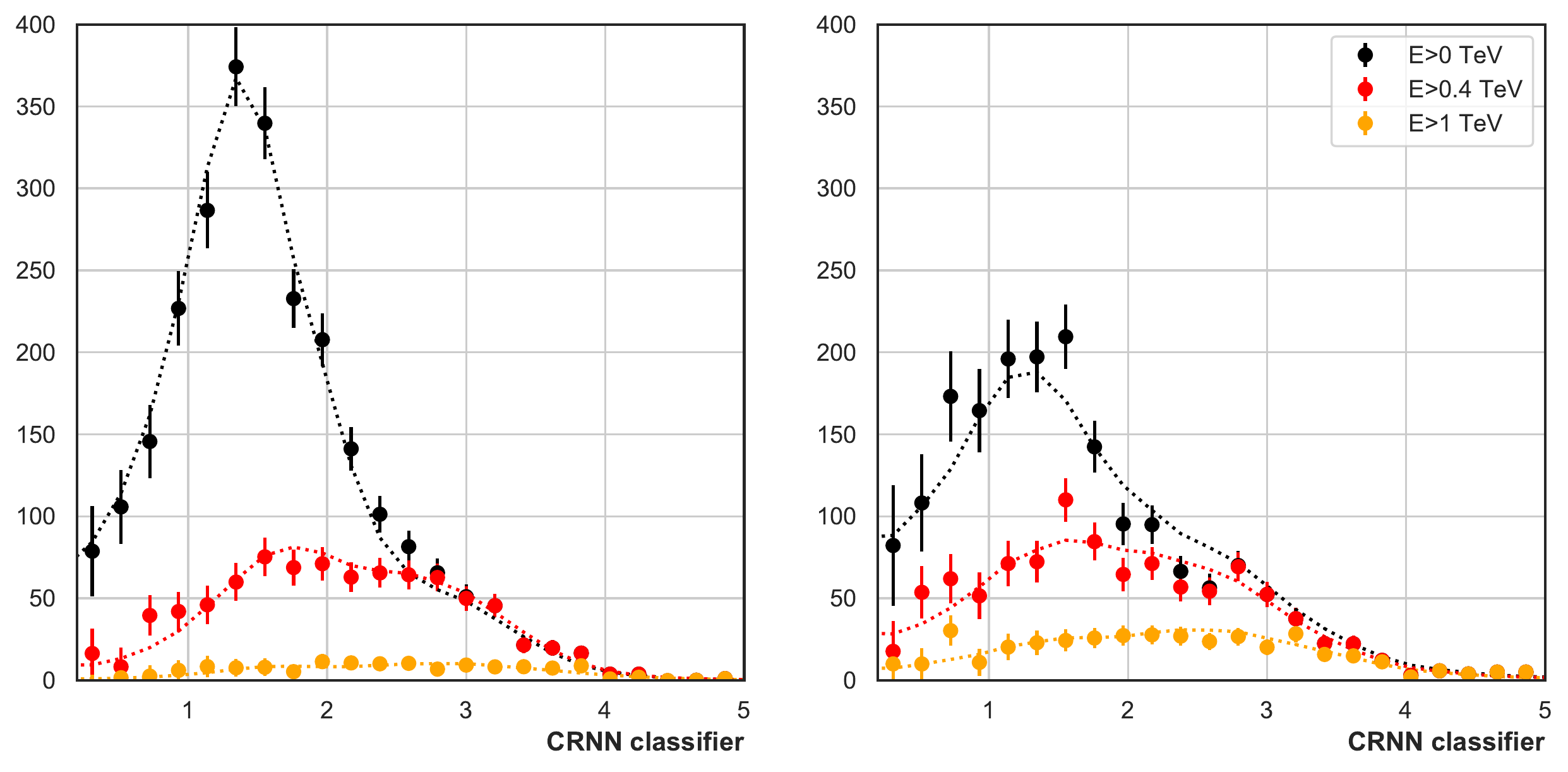}

\caption{Comparison of the CRNN classifier distribution with three energy thresholds obtained in the analysis of PKS\,2155-304 (left) and HESS J1745-290 (right), data is compared with to Monte Carlo simulations re-weighted to match the source spectrum index with 100 and 200\,MHz NSB level respectively, and scaled to number of excess events in each energy bin.}
\label{fig:SourceResult}
\end{center}
\end{figure*}

\subsection*{PKS\,2155-304}

The first source tested was the well known BL Lac object PKS\,2155-304 observed throughout the operation of the H.E.S.S. instrument (e.g. \cite{PKS2155HESS,PKS2155HESS2}). 
A sample of around 15 hours of observation with zenith angle of close to 20$^\circ$ and optical efficiency similar to that in the MC simulations was selected from the non-flaring periods of PKS\,2155-304. This dataset contains a similar number of more than 1300 excess events and takes place over a relatively diverse set of observing conditions. Figure \ref{fig:SignificanceMap} (left) shows the resultant significance map (created using the \emph{gammapy} software package \cite{gammapy}) of this region, clearly showing a strong source at a position consistent with the catalogue position of PKS 2155-304. Figure \ref{fig:SignificanceMap} (right) shows the distribution of significance from non-source pixels, which is well fit by a Gaussian with a mean of 0.14 and a width of 1.07, quite consistent with the expectation for well normalised background in signal free regions (mean of 0, width of 1).

Table \ref{tab:stats} shows the detection statistics for the dataset using the \emph{reflected background} \cite{bergeBackground} to estimate the residual background contamination in the source region. For all network configurations tested a similar number of excess gamma-ray events are detected due to the cut being made on the expected gamma-ray efficiency. As expected the RNN-based networks show a reduction in the estimated level of background contamination. However as in this case as there is some small variation in the number of excess events it is fairer to make comparisons of the signal to background ratio (S/B) i.e. the number of excess events divided by estimated background contamination. 
In both the 60\% and 80\% cut set the Hillas-RNN shows an improvement in S/B of around 5-10\%, while the CRNN shows an improvement of around 20\% over the mean scaled network. This improvement in background rejection does not translate into large increases in source significance due to the extremely bright source being investigated.

Figure \ref{fig:ClassDist} (left) shows the distribution of the CRNN classification parameter obtained from this datasets in comparison with the results of MC simulations (at 100 MHz NSB rate) re-weighted to a spectral index of -3.4. In this case the data distribution provides an excellent match to the Monte Carlo expectation, demonstrating a stable behaviour of the classifier on strong, steep spectrum sources.

\begin{table*}
\begin{center}
\scalebox{0.9}{
\begin{tabular}{c|c|c|c|c|c|c||c|c|c|c|c|}
\cline{3-12}
\multicolumn{2}{ c }{} &\multicolumn{5}{ |c|| }{\emph{80\% Gamma-ray Efficiency}} &  \multicolumn{5}{ c| }{\emph{60\% Gamma-ray Efficiency}} \\ \cline{2-12}
& \textbf{Network}& \textbf{N$_{\rm{ON}}$} & \textbf{$\alpha$ N$_{\rm{OFF}}$} & \textbf{Excess} & \textbf{S/B} & \textbf{$\sigma$}  & 
\textbf{N$_{\rm{ON}}$} & \textbf{$\alpha$ N$_{\rm{OFF}}$} & \textbf{Excess} & \textbf{S/B} & \textbf{$\sigma$} \\ \cline{1-12} 

\multicolumn{1}{ |c| }{}& \textit{MSC NN} & 2602 & 560.4 & 2041.6 & \textbf{ 3.64 } & \textbf{ 57.7 }
& 1841 & 288.7 & 1552.3 & \textbf{ 5.38 } & \textbf{ 55.7 }  \\ \cline{2-12}
\multicolumn{1}{ |c| }{\textbf{PKS 2155-304} }& \textit{Hillas RNN} & 2590 & 529.4 & 2060.6 & \textbf{ 3.89 } & \textbf{ 59.1 }
& 1825 & 268.7 & 1556.3 & \textbf{ 5.79 } & \textbf{ 56.7 } \\ \cline{2-12}
\multicolumn{1}{ |c| }{(quiescent)}& \textit{CRNN} & 2634 & 477.1 & 2156.9 & \textbf{ 4.52 } & \textbf{ 62.8 }
& 1904 & 248.5 & 1655.5 & \textbf{ 6.66 } & \textbf{ 60.4 }\\ \cline{1-12} 

\multicolumn{1}{ |c| }{}& \textit{MSC NN} & 3071 & 1553.4 & 1517.6 & \textbf{ 0.98 } & \textbf{ 31.5 }
& 2068 & 844.2 & 1223.8 & \textbf{ 1.45 } & \textbf{ 32.8 }\\ \cline{2-12}
\multicolumn{1}{ |c| }{\textbf{HESS J1745-290}}& \textit{Hillas RNN} & 2813 & 1327.1 & 1485.9 & \textbf{ 1.12 } & \textbf{ 32.9 }
& 1906 & 716.5 & 1189.5 & \textbf{ 1.66 } & \textbf{ 33.9 } \\ \cline{2-12}
\multicolumn{1}{ |c| }{}& \textit{CRNN} & 2968 & 1320.2 & 1647.8 & \textbf{ 1.25 } & \textbf{ 36.0 }
& 2030 & 693.0 & 1337.0 & \textbf{ 1.93 } & \textbf{ 37.8 }\\ \cline{1-12}
\end{tabular}
}
\end{center}

\caption{Detection statistics for the two run lists tested with background cuts tuned to retain 80\% and 60\% of the gamma-ray events at all energies.}
\label{tab:stats}

\end{table*}

\subsection*{HESS\,J1745-290}

The second case studied was the Galactic Centre point source HESS\,J1745-290 \cite{SgrA} commonly associated with the supermassive black hole Sagittarius A*. In this case a selection of data was made from 2004 to 2008 datasets, resulting a total of around 30 hours of observations. This field of view represents a rather different analysis proposition to PKS\,2155-304. Firstly the spectrum of this source is comparatively hard, with a spectral index of -2.1 and a cut-off at around 14 TeV. In addition to this the level of nightsky background in this region is significantly higher at around 200\,MHz in comparison with the approximately 60\,MHz in the PKS 2155-304 observations. For this dataset an improvement in S/B of around 15\% is seen for the Hillas-RNN over the mean scaled NN and around 25\% in the CRNN.

The CRNN classifier distribution shown in figure \ref{fig:ClassDist} (right) again shows an excellent match to the MC simulations (with an NSB level of 200 MHz) re-weighted to the source spectrum. Again this demonstrated the stable behaviour of the network even with diverse observation conditions and higher NSB levels, although clearly care must be taken to choose the correct NSB level in the simulations.

\section*{Discussion}

In this paper we have demonstrated the potential sensitivity gains available to imaging atmospheric Cherenkov telescopes by using the latest generation of machine-learning tools for background rejection and for the first time demonstrated a successful application of this scheme to data from the H.E.S.S. gamma-ray observatory. Applications of the convolutional-recurrent neural network to Monte Carlo air shower simulations and real data show an improvement in background rejection power of around 20-25\% over the use of mean-scaled parameters typically used in previous background rejection implementations. 

Although this does not match the even stronger performance gains predicted from simulated events (20\% and 60\% for the Hillas RNN and CRNN respectively) this mismatch could be caused by several factors. First the presence of cosmic ray electrons which are present in the data (e.g. \cite{HESSelectrons}) is not accounted for in the simulation predictions. These electron induced air showers develop almost identically to gamma-ray induced air showers and are often considered to represent an irreducible background in IACT data which becomes more and more important as the hadron rejection power improves. 

Secondly the network training is performed using simulated protons as the background events, however significant systematic uncertainties exist in the modelling of hadronic interactions in this energy range \cite{ParsonsHadron}. This behavioural uncertainty could result in a reduced performance when applying the trained networks to data due to incorrectly reproducing features  within the air shower. However, due to the "black box" nature of network behaviour it is difficult to identify any features that do not match between data and simulations.

This improvement is in line with the performance of that of goodness of fit cuts from image template based event reconstruction (e.g. \cite{deNaurois2009}). The reproduction of the sensitivity of goodness of fit based cuts is to be expected in the case of gamma-rays where the air showers behave in relatively predictable way and the images seen in the individual telescopes are strongly correlated.
It is important to take note of the sensitivity of the network performance to different observing conditions and that care must be taken when to ensure that particularly the night-sky background level of the simulations matches that of the data to which it is being compared. This strong sensitivity to NSB level could potentially be lowered by careful preprocessing and denoising of the image, however it is possible that a run-wise-simulation scheme (e.g. \cite{RWS}) may be required to ensure the lowest possible systematic uncertainties if such a scheme is deployed.

\section*{Conclusion}

Although no significant performance gains are seen in background rejection power over the current state of the art goodness of fit based background rejection, use of this machine learning scheme does add some benefits. Firstly, the systematic uncertainties of this method, while likely as large, are different from the goodness of fit based approach. Thus allowing evaluation of the systematic uncertainties of analysis at the limits of the instrumental threshold. Secondly, the goodness of fit based approach relies on comparing shower images to a mean expected image template, limiting it's usefulness in the classification of particle species which produce large shower-to-shower fluctuations (such as protons or heavier nuclei). However, the training step of the RNNs naturally includes these fluctuations, meaning the RNNs may also be extremely useful in measuring the mass composition of hadrons in IACT data.

Applications of this neural network structure are not limited to event classification in IACTs and with some modification could be applied to regression problems such as direction and energy reconstruction. 

\section*{Acknowledgements}

The authors would like to thank the H.E.S.S. Collaboration for allowing the use of H.E.S.S. data and simulations in this publications, as well as providing useful discussions and input to the paper.


\bibliographystyle{number_cite}

\bibliography{RNN}

\end{document}